\documentclass[polymers,article,submit,pdftex,moreauthors]{Definitions/mdpi} 
\renewcommand{\linenumbers}{}
\firstpage{1} 
\pubvolume{1}
\issuenum{1}
\articlenumber{0}
\pubyear{2022}
\copyrightyear{2022}
\datereceived{} 
\dateaccepted{} 
\datepublished{} 
\hreflink{https://doi.org/} % If needed use \linebreak
\Title{Self-attractive semiflexible polymers under an external force field}

\TitleCitation{Self-attractive semiflexible polymers under an external
  force field}

\Author{Antonio Lamura $^{1}$\orcidA{}}

\AuthorNames{Antonio Lamura}

\AuthorCitation{Lamura, A.}
\address{%
  $^{1}$ \quad Istituto Applicazioni Calcolo, Consiglio Nazionale delle Ricerche
  (CNR), Via Amendola 122/D, 70126 Bari, Italy; antonio.lamura@cnr.it}

\abstract{The dynamical response of a tethered semiflexible polymer
with self-attractive
interactions and subjected to an external force field is numerically
investigated by varying stiffness and self-interaction strength.
The chain is confined in two spatial dimensions
and placed in contact with a heat bath described by the Brownian multiparticle
collision method. For strong self-attraction the equilibrium conformations
range from compact structures to double-stranded chains, and to rods
when increasing the stiffness. Under the external field 
at small rigidities, the initial close-packed chain
is continuously unwound by the force before being completely elongated.
For double-stranded conformations the transition from the folded state to
the open one
is sharp being steeper for larger stiffnesses.
The discontinuity in the transition
appears in the force-extension relation as
well as in the probability distribution function of the gyration radius.
The relative deformation with respect to the equilibrium case
along the direction normal to the force is found to decay as the inverse
of the applied force. 
}

\keyword{mesoscale simulations; nonequilibrium simulations; polymer dynamics} 

\begin{document}

%%%%%%%%%%%%%%%%%%%%%%%%%%%%%%%%%%%%%%%%%%
\section{Introduction}

The study of single polymers such as, for example, DNA, filamentous actin,
and microtubules under various flow conditions, has helped in understanding
their dynamical and conformational properties \cite{shaq:05}.
The first investigations of the flow behavior of single DNA filaments
\cite{perk:95} opened the way to a large variety of flow experiments
which provided insight into the  mechanisms regulating the dynamics. Several
computational models have been studied
which reveal to be very useful in understanding such systems.
Single polymer studies give the chance of directly observing the microscopic
conformations of individual chains close to equilibrium or
under flow conditions, thus accessing non-equilibrium conformations.

In the case of biological filaments, their stiffness
is closely related to their functions. For example, the rigidity of actin
filaments is responsible for the mechanical properties of the
cytoskeleton, and DNA is able to pack in the genome or inside a virus
capsid thanks to its persistence length. Several works
have investigated the equilibrium properties of semiflexible polymers
\cite{wilh:96,goet:96,harn:96,ever:99,samu:02,lego:02,wink:03,petr:06}. 
The development of spectroscopic techniques and
fluorescence microscopy provided insight into their non-equilibrium properties
(for reviews see, e. g., Refs.~\cite{schr:18,yeou:22}).
Theoretical 
\cite{bird:87,oett:96,wink:06_1,munk:06,wink:10}
and computational 
\cite{hur:00,hsie:04,liu:04,send:08,he:09,
zhan:09,lang:14,niko:17,lamu:19,shee:21,lamu:21,ande:22} 
studies helped in revealing
and understanding novel dynamical, conformational and rheological properties.

Among others, the worm-like chain model \cite{krat:49}
proved to be accurate to describe
the mechanical response of semiflexible polymers under specific conditions.
Indeed, the main
limitations of this model come from neglecting excluded volume effects
and self-interactions between different polymer parts. The former are relevant
especially in two dimensions leading, for example, to segregation of
polymers \cite{witt:10}. The latter interactions, that are not relevant
for strong applied fields or far from the folding temperature, are crucial
for semiflexible chains with monomer-monomer interactions, such as
poly(ethylene oxide) (PEO), DNA \cite{baum:00} or RNA \cite{gerl:03}
in poor-solvent condition \cite{doi:86}.
Short-range attractive interactions lead to a large variety
of conformations due to the competition of polymer stretching and
collapse \cite{aust:17}.
Previous experimental \cite{haup:02,guna:07}, theoretical and numerical
\cite{rosa:03,tkac:04,kuma:05,knel:05,rosa:06,kuma:07,cifr:07,kapr:09,gutt:09}
studies have found that the mechanical
response of self-interacting
semiflexible polymers to an external stretching is very complicated.
These investigations considered either chains
with one of its ends grafted and the other
one pulled by a force, or chains with both ends pulled away in opposite
directions.

A large majority of studies has been
performed in three dimensions but
addressing the comprehension of stretched self-interacting polymers
in two dimensions is also interesting
for two main reasons: Excluded-volume effects are relevant and
hydrodynamic interactions can be
neglected in the case of polymers strongly adsorbed on surfaces since
the overall dynamics is dominated by the polymer-substrate
interaction \cite{maie:02}. 
Two-dimensional realizations of these systems can, for example,
be provided by DNA strongly adsorbed on a surface with one grafted end.
Under these conditions the stretching of biopolymers
is observed
in systems with separation of biomolecules
via nanochannels \cite{kim:11,reis:12}.
The effects of a uniform force field on two-dimensional
semiflexible polymers have been
considered both in experimental \cite{maie:02} and numerical
\cite{lamu:01.1,lamu:12} studies but neglecting self-attractions among
monomers.

So far a systematic study of polymers under poor-solvent condition
in an external field is lacking.
In the present work, the dynamical and conformational properties
of a semiflexible filament, tethered by one of its ends and
subjected to an external force field
in two spatial dimensions, are numerically
investigated. The polymer is modeled as a self-avoiding worm-like chain
with self-attraction among beads.
Hydrodynamics is neglected since it is assumed that local polymer friction
is uniquely fixed by its interaction with the adsorbing surface.
For this reason the polymer is taken to be in contact
with a Brownian heat bath. This is implemented by adopting the Brownian version
\cite{ripo:07} of the multiparticle collision dynamics \cite{kapr:08,gomp:09}.
By varying stiffness
and self-interaction strength, different equilibrium conformations are
found. For strong mutual attraction
and relative low stiffness, the structure is compact.
Increasing the chain rigidity promotes
the formations of folded strands.
The mechanical response of the polymer to the applied force 
depends on the equilibrium structure.
At small rigidities the initial close-packed chain is continuously
unwound by the external force field. The polymer shows bistable conformations
before being completely elongated.
When double-stranded chains form,
a ``first-order''-like phase transition
to the open conformation is observed in the
force-extension curve.  
Polymer configurations are characterized by considering the gyration
tensor: It is found that the relative deformation
with respect to the equilibrium case along the direction normal
to the force, decays as the inverse of the applied force.

The numerical model for the polymer and the Brownian heat bath are
illustrated in  Section \ref{sec:model}. The results for the equilibrium
conformations and the dynamic behavior are reported in
Section \ref{sec:results}.
Finally, in Section \ref{sec:conclusions} the main findings
of this study are discussed drawing some conclusions.

%%%%%%%%%%%%%%%%%%%%%%%%%%%%%%%%%%%%%%%%%%
\section{Model and Method} \label{sec:model}

A linear chain of length $L$, made 
of $N+1$ beads of mass $M$, is considered in two spatial dimensions.
Internal forces acting on beads are due to a potential  
which accounts for different contributions.
Connected beads interact via the harmonic potential 
\begin{equation}\label{bond}
U_{bond}=\frac{\kappa_h}{2} \sum_{i=1}^{N}
(|{\bf r}_{i+1}-{\bf r}_{i}|-r_0)^2 ,
\end{equation}
where ${\bf r}_i=(x_i,y_i)$ denotes the position vector of the $i-$th bead
($i=1,\ldots,N+1$), $r_0$ is the average bond length, and
the elasticity is controlled by $\kappa_h$.
The parameter $\kappa_h$ is chosen in order to preserve on the average
the total contour length $L= N r_0$ .
Chain stiffness of the polymer is introduced via the bending potential
\begin{equation}
U_{bend}=\kappa \sum_{i=1}^{N-1} (1-\cos \varphi_{i})
\label{bend}
\end{equation}
where $\kappa$ is
the bending rigidity and $\varphi_{i}$ is the angle between
two consecutive bond vectors. 
Non-bonded pair interactions are
modeled by the Lennard-Jones potential
\begin{equation}
U_{LJ} = 4 \epsilon \sum_{i=1}^{N-1} \sum_{j=i+2}^{N+1}
\Big [ \Big(\frac{\sigma}{r_{i,j}}\Big)^{12}
-\Big(\frac{\sigma}{r_{i,j}}\Big)^{6} \Big] ,
\label{rep_pot}
\end{equation}
where $r_{i,j}$ is the distance between two non-consecutive beads.
A strongly attractive regime corresponds to energies $\epsilon > k_B T$, which
determine compact structures. In the opposite limit $\epsilon < k_B T$ of weak
self-attraction, swollen chain configurations can be observed. Here
$k_B T$ is the thermal energy, $T$ is the temperature, and $k_B$ is
Boltzmann's constant.
The parameters $\kappa$ and $\epsilon$ are varied in the present study,
keeping fixed the temperature,
to obtain different equilibrium conformations as later shown.
In the following, for the sake of clarity, chain stiffness is characterized
in terms of the length $L_p=2 \kappa r_0/ k_B T$.
In the worm-like chain limit, when the Lennard-Jones potential $U_{LJ}$ is
negligible,
this length corresponds to the polymer persistence
length \cite{wink:94}. However, in the present model
this is not strictly true due to the coexistence of different
length and energy scales \cite{bind:10}.
Finally, in order to consider external stretching of the chain,
a constant force $F$ acts on every bead of the polymer.
This force is directed along the $x$-direction of the Cartesian reference frame
and corresponds to an external potential given by
\begin{equation}\label{potforce}
U_{ext}=-F\sum_{i=1}^{N+1} x_{i} .
\end{equation}
The external field could be a gravitational or uniform flow field.
Newton's equations of motion of beads are integrated by
the velocity-Verlet algorithm with time step $\Delta t_p$
\cite{swop:82,alle:87}.

The chain is coupled to a Brownian heat bath which is
implemented by using the Brownian multiparticle collision (B-MPC) method
\cite{ripo:07,gomp:09,kiku:03} without taking into account hydrodynamic
interactions. Here we adopt the computationally efficient
version proposed in Ref.~\cite{ripo:07}.
In this algorithm every bead undergoes
stochastic collisions with a 
virtual particle of mass $M$
to simulate the interaction with a
fluid volume surrounding the bead. The momenta of such phantom
particles are Maxwell-Boltzmann distributed with
variance $M k_B T$ and zero mean.
The collision process
is implemented via the stochastic rotation dynamics
of the MPC method \cite{ihle:01,lamu:01,gomp:09}. This corresponds to randomly
rotate the
relative velocity of a polymer bead, with respect to the
center-of-mass velocity of the bead and its related phantom
particle, by angles $\pm \alpha$. 
Collisions occur at time intervals $\Delta t$ being $ \Delta t > \Delta t_p$. 

Simulations are performed with the choices $\alpha=130^{o}$,
$\Delta t=0.1 t_u$, with time unit $t_u=\sqrt{m r_0^2/(k_B T)}$,
$M =5 m$, $\kappa_h r_0^2/(k_B T)=10^4$, 
$\sigma=r_0$, $N=50$,
and $\Delta t_p=10^{-2} \Delta t$. The value of
$\kappa_h$ ensures that the polymer length $L$ is constant within
$1\%$ for all systems.
 
%%%%%%%%%%%%%%%%%%%%%%%%%%%%%%%%%%%%%%%%%%
\section{Numerical results} \label{sec:results}

Polymers are initialized with beads
randomly aligned along the $x$-direction and allowed to equilibrate.
The position ${\bf r}_1$ of the first bead is fixed
at the origin $(0,0)$ of the Cartesian reference frame
while no orientation is enforced for the first bond.
When taking into account
the action of the uniform force field, simulations are started
from the equilibrium configurations of chains and run until
reaching steady states during which average quantities are computed.
We consider semiflexible polymers with values of the bending rigidity
$\kappa$ such that $0.1 \leq L_p/L \leq 2$, and interaction energies
$\epsilon / k_B T = 0.25, 2$.

\subsection{Equilibrium polymer conformations} \label{sec:conf}

\begin{figure}[H]
\includegraphics*[width=.7\textwidth,angle=0]{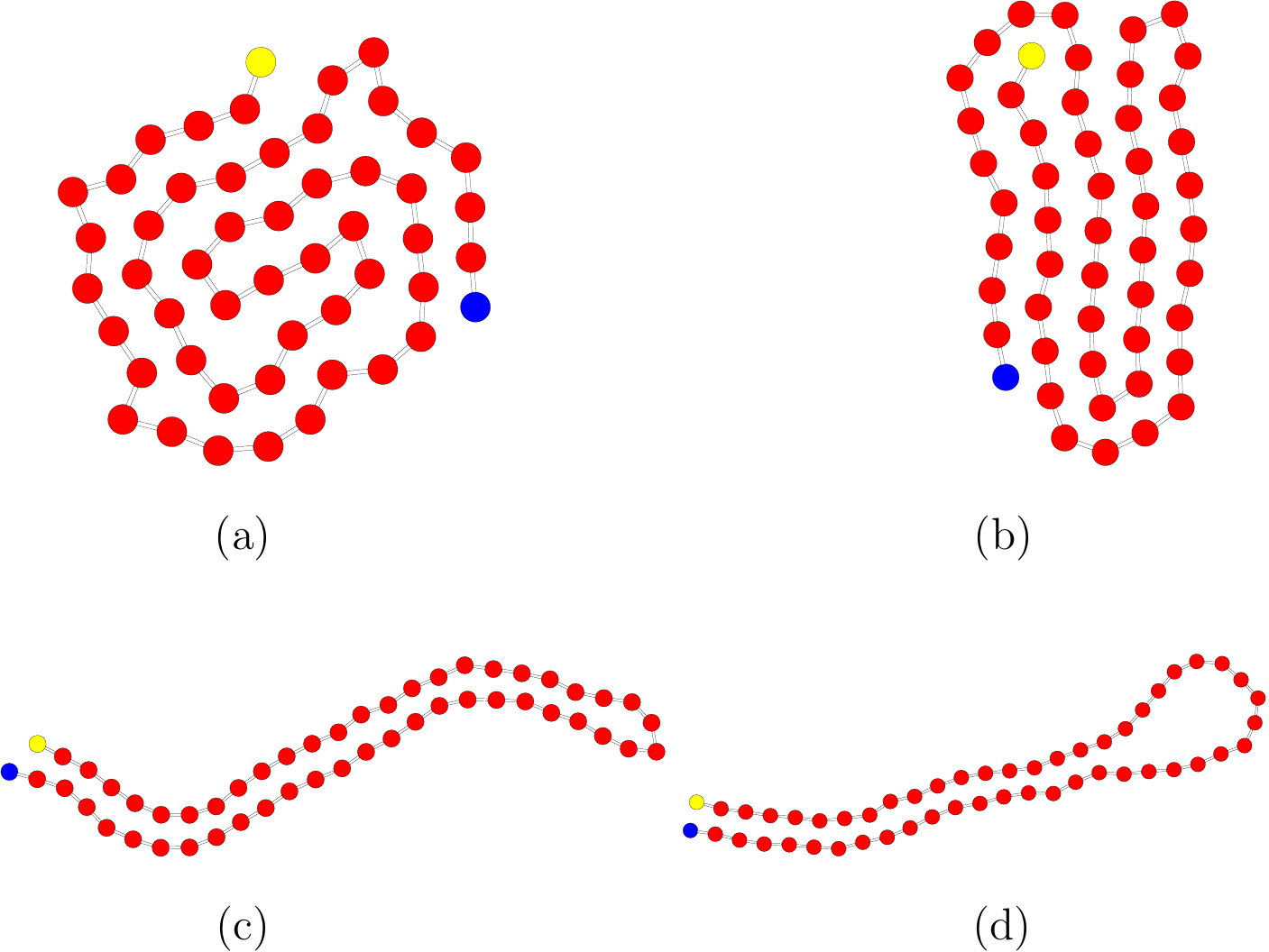}
\caption{Equilibrium polymer conformations for
$L_p/L=0.1 (a), 0.2 (b), 0.4 (c), 0.8 (d)$
with $\epsilon / (k_B T) = 2$. Blue and yellow beads denote the first and last
ones, respectively. Polymer beads and bonds are not in scale to allow
a better visualization.
\label{fig:conf0}
}
\end{figure}

In this Section the equilibrium properties of polymers are obtained and
characterized by varying the bending rigidity and the interaction energy.
When considering the value $\epsilon / k_B T = 0.25$, non-bonded
interactions are  negligibly small and the model
corresponds
to the worm-like chain model \cite{krat:49}, as shown in the following.
In this case the filament
assumes a swollen configuration with spatial correlations, in the direction
of the chain tangent, on a length scale given by the persistence length.
A different scenario occurs when non-bonded interactions become relevant.
Equilibrium configurations for $\epsilon / k_B T = 2$ and different values
of the length $L_p$ are shown in Figure \ref{fig:conf0}.
For the smallest
value $L_p/L=0.1$ (Figure \ref{fig:conf0} (a)) the chain has a globule
structure which is very compact.
Increasing the stiffness promotes the formation of folded bundles.
A configuration
with five rod-like strands is shown
in Figure \ref{fig:conf0} (b)
for $L_p/L=0.2$. The energy penalty, which is proportional to
$(1-\cos \varphi)$ and
increases with the bending angle $\varphi$ at turning points,
is compensated by the
energy gain from bead-bead attractions. 
The number of strands diminishes when increasing $\kappa$.
A structure formed by two
facing strands is observed at $L_p/L=0.4$ (Figure \ref{fig:conf0} (c)).
For this value of $L_p$ the average bending energy diminishes
since the number of turning points reduces, and
the average Lennard-Jones energy increases.
A further increase in chain stiffness
induces the formation of hairpin conformations (Figure \ref{fig:conf0} (d)
for $L_p/L=0.8$). This causes a second rise in the bending energy
whose energetic penalty can still be compensated
by the mutual attraction between
monomers. Finally, at $L_p \simeq L$ the polymer cannot sustain any closed
configuration and a rod-like
structure is observed for values $L_p \gtrsim L$.
In this latter range the average value of $U_{LJ}$ exhibits a sharp
increase while the average bending energy decreases.

\begin{figure}[H]
\includegraphics*[width=.5\textwidth,angle=0]{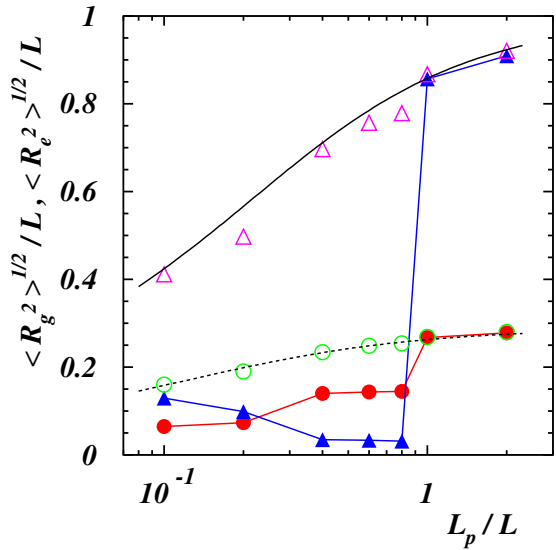}
\caption{Root-mean-square end-to-end distance $R_e$
of a polymer in absence of an
external force for
 $\epsilon / (k_B T) = 0.25$ (purple open triangles),
 $2$ (blue filled triangles), and gyration radius $R_g$ for
 $\epsilon / (k_B T) = 0.25$ (green open circles),
 $2$ (red filled circles).
 The full and dashed black lines correspond to the analytical predictions
(\ref{eq:re}) and (\ref{eq:rg}), respectively, in the case of continuous
semiflexible polymers \cite{wink:94}.
\label{fig:raggi}
}
\end{figure}

In order to characterize the conformations of chains, it is useful
to consider the root-mean-square values of the 
end-to-end distance $\langle R_e^2 \rangle^{1/2}$, where
$R_e=|{\bf r}_{N+1}-{\bf r}_1|$, and of the gyration radius
$\langle R_g^2 \rangle^{1/2}$. By computing the gyration tensor
\begin{equation}
G_{\alpha\beta}=\frac{1}{N+1}\sum_{i=1}^{N+1}  \Delta r_{i,\alpha}
\Delta r_{i,\beta} ,
\label{eq:tens}
\end{equation}
where $\Delta r_{i,\alpha}$ is the position of the $i$-th monomer
in the center-of-mass reference frame of the chain and the Greek index
denotes the Cartesian component, the gyration radius can be obtained as 
$R_g^2=\sum_{\alpha} G_{\alpha\alpha}$.
The computed values of $\langle R_e^2 \rangle^{1/2}$ and $\langle R_g^2 \rangle^{1/2}$ for
the two values of $\epsilon$ as functions of the dimensionless length $L_p/L$
are presented in Figure \ref{fig:raggi}.
For the smallest value
$\epsilon / k_B T=0.25$ the numerical results show a quantitative agreement
with the theoretical predictions for
a continuous semiflexible chain \cite{wink:94}
\begin{eqnarray}
\langle R_e^2\rangle &=& 2 L_p L \Big [ 1 -\frac{L_p}{L}\Big(1-e^{-L/L_p}\Big) \Big ] ,
\label{eq:re}\\
\langle R_g^2\rangle &=& L_p L \Big [ \frac{1}{3}-\frac{L_p}{L}
+2 \Big(\frac{L_p}{L} \Big)^2
-2 \Big(\frac{L_p}{L} \Big)^3\Big(1-e^{-L/L_p}\Big) \Big ] .
\label{eq:rg}
\end{eqnarray}
This confirms that 
the self-interaction energy is negligible for this choice of $\epsilon$ and
the polymer behaves as a worm-like chain. 
The behavior is different for the highest value of the energy $\epsilon$.
The end-to-end distance is smaller than in the previous case and decreases
to reach its minimum value when the chain consists of two strands folded on
each other
($0.4 \lesssim L_p/L \lesssim 0.8$). Then, $\langle R_e^2 \rangle^{1/2}$
jumps to values comparable to those of semiflexible polymers
at $L_p/L \simeq 1$. 
The average gyration radius $\langle R_g^2 \rangle^{1/2}$ is at a minimum
when compact conformations are observed ($L_p/L \lesssim 0.2$), then increases to a value
which remains constant as long as the chain consists of two strands, and finally
reaches the equilibrium values of worm-like chains when the polymer
assumes a rod-like structure.

\begin{figure}[H]
\includegraphics*[width=.5\textwidth,angle=0]{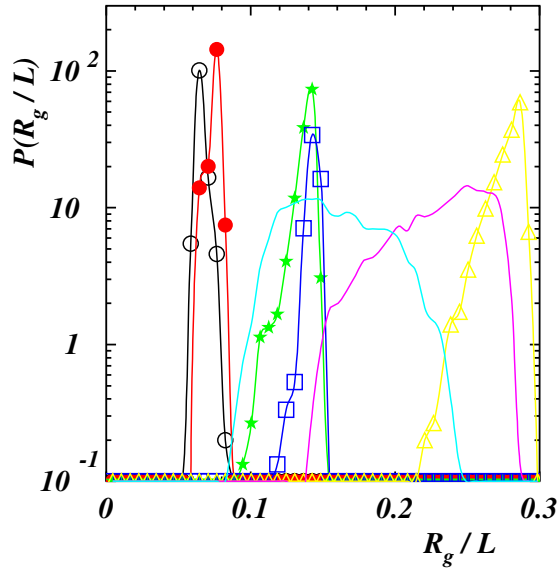}
\caption{Normalized probability distribution function of the
gyration radius $R_g$ in absence of the external force for
$L_p/L=0.1$ (black open circles), $0.2$ (red filled circles),
$0.4$ (green  filled stars), $0.8$ (blue open squares),
$2$ (yellow open triangles) with $\epsilon / (k_B T) = 2$,
and for 
$L_p/L=0.1$ (cyan line), $0.4$ (purple line) with $\epsilon / (k_B T) = 0.25$.
\label{fig:pdfrg0}
}
\end{figure}

Normalized probability distribution functions (PDFs) of the polymer
gyration radius $R_g$ are depicted in Figure \ref{fig:pdfrg0} under equilibrium
conditions.
For the highest value of the interaction energy $\epsilon$,
when the polymer is compact,
curves are very narrow corresponding to the fact that the chain
global conformation does not change significantly in time. The two curves with
$L_p/L=0.1, 0.2$ 
almost overlap with peaks located at $R_g/L \simeq 0.07$.
When considering double-stranded chains,
the curve at $L_p/L=0.4$ is broader since the chain
fluctuates along its length. The peak is at
$R_g/L \simeq 0.14$ as in the case with $L_p/L=0.8$ where
the PDF is narrower since the structure is quite rigid.
Finally, when the polymer assumes a rod-like conformation ($L_p/L=2$),
the position of the PDF peak moves to $R_g/L \simeq 0.28$.
For a comparison two PDFs in the case of weak self-attraction
($\epsilon / k_B T=0.25$) are also presented in the figure.
Curves are broader than in the previous case due to the fact that chains
are more prone to fluctuate since the mutual attraction is negligible.
The peaks
are located at larger values of $R_g$ with respect to the case with
$\epsilon / k_B T=2$, for the same stiffness, corresponding to more
elongated structures. 

\subsection{Polymer stretching in uniform force field} \label{sec:force}

When the polymer is subject to the external force, it is stretched along
the direction of the force. 
In order to characterize the elongation of the chain, the
average deficit length-ratio $\delta=1-\langle x_{N+1} \rangle/L$
as a function of the
applied force is considered. $\langle x_{N+1} \rangle$
is the average
extension of the chain along the force direction computed
as the average value of the $x$-component of the end-to-end vector
${\bf R_e}={\bf r}_{N+1}-{\bf r}_1$.
When self-attraction is negligible, 
in the limit $|x_{N+1}| \rightarrow L$ it results
\cite{mark:95,maie:02}
\begin{equation}
\delta \sim \left (\frac{1}{F_2} \right )^{1/2} 
\label{sf}
\end{equation}
with $F_2=N F L_p / (k_B T)$.
For quite strong force fields or very small bending rigidities
the behavior does not depend on the stiffness
and is given by \cite{lamu:01.1,liva:03,rosa:03}
\begin{equation}
\delta \sim \frac{1}{F_1} 
\label{f}
\end{equation}
where $F_1=N F r_0 / (k_B T)$, as for flexible chains \cite{rubi:03}.

\begin{figure}[H]
\includegraphics*[width=.9\textwidth,angle=0]{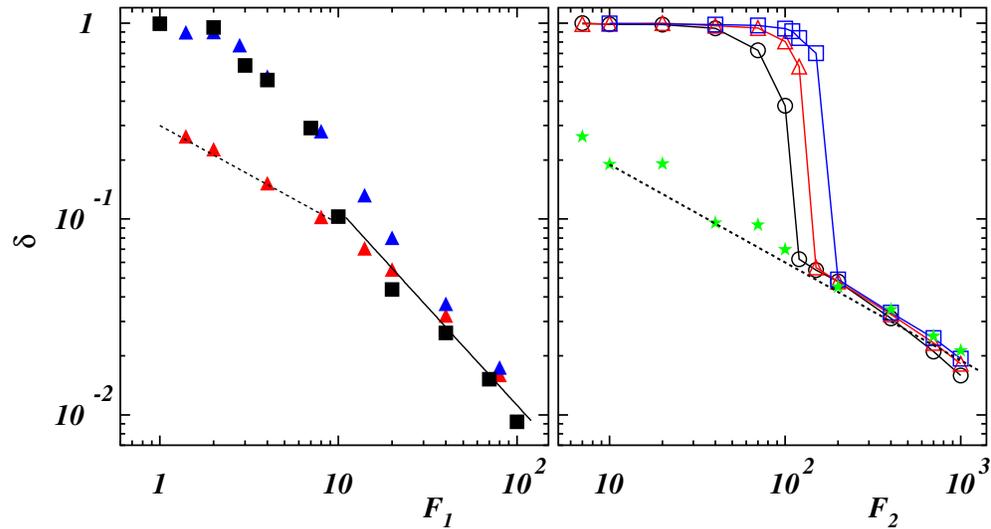}
\caption{(Left panel) Mean deficit length-ratio along the direction of the external force
as a function of the dimensionless force $F_1=N F r_0 / (k_B T)$
for $L_p/L=0.1$ (blue filled triangles),
$0.2$ (black  filled squares) with $\epsilon / (k_B T) = 2$,
and for $L_p/L=0.1$ (red filled triangles)
with $\epsilon / (k_B T) = 0.25$. The dashed and full
lines have slopes $-1/2$ and $-1$, respectively.
(Right panel) Mean deficit length-ratio
along the direction of the external force
as a function of the dimensionless force $F_2=N F L_p / (k_B T)$
for $L_p/L=0.4$ (black open circles),
$0.6$ (red open triangles),
$0.8$ (blue open squares), $2$ (green filled stars)
with $\epsilon / (k_B T) = 2$. The dashed
line has slope $-1/2$.
\label{fig:deficit}
}
\end{figure}

\begin{figure}[H]
\includegraphics*[width=.9\textwidth,angle=0]{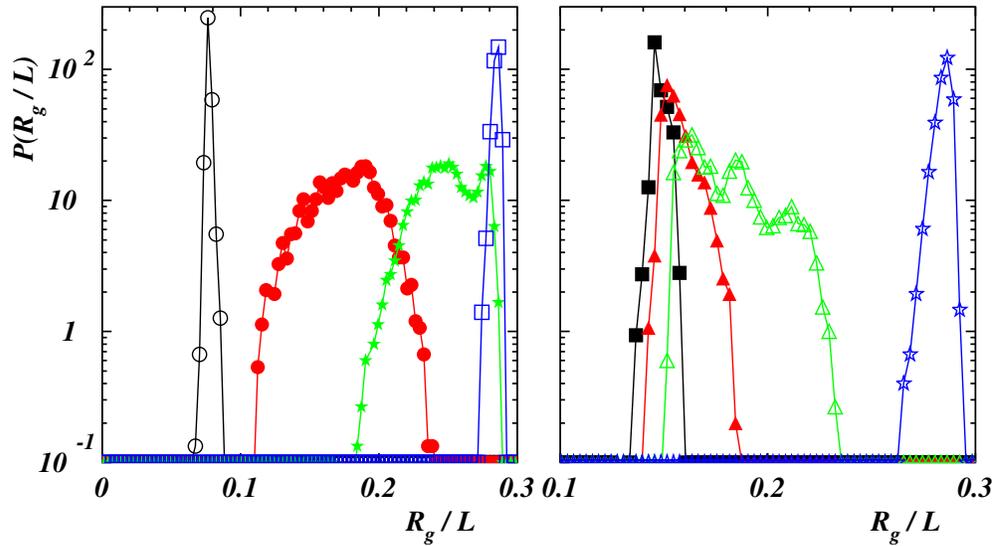}
\caption{(Left panel)
Normalized probability distribution function of the
gyration radius $R_g$ for
$F_1=N F r_0 / (k_B T)=1$ (black open circles), $4$ (red filled circles),
$7$ (green  filled stars), $20$ (blue open squares)
with $L_p/L=0.2$ and $\epsilon / (k_B T) = 2$.
(Right panel)
Normalized probability distribution function of the
gyration radius $R_g$ for
$F_2=N F L_p / (k_B T)=100$ (black filled squares),
$120$ (red filled triangles),
$150$ (green open triangles), $200$ (blue filled stars)
with $L_p/L=0.8$ and $\epsilon / (k_B T) = 2$.
\label{fig:pdfrg}
}
\end{figure}

Different behaviors
can be expected for self-interacting semiflexible polymers.
When the filament is pulled at one end by
a constant force, a sharp transition appears in the force
{\it vs.} elongation curves \cite{rosa:03} whose sharpness is enhanced
by bending rigidity \cite{rosa:06}.
Simulations results of the present model 
are illustrated in Figure \ref{fig:deficit} as functions of applied force
for different values of the ratio $L_p/L$.
In case of $L_p/L \leq 0.2$, corresponding to compact initial states
(see Figure \ref{fig:conf0}), data collapse is obtained when plotting values
of $\delta$
as functions of the dimensionless force $F_1=N F r_0 / (k_B T)$
(left panel of Figure \ref{fig:deficit}). 
The initial structure is tilted in the
direction of the force and only slightly deformed as long as $F_1 \lesssim 1$.
This can also be appreciated when considering the normalized PDFs
of the gyration radius: In the case with $L_p/L=0.2$ and $F_1=1$, the
PDF exhibits a narrow peak
(see Figure \ref{fig:pdfrg} (left panel)).
By increasing the force, the extension increases smoothly since
the globule is partially unwound,
similarly to what holds for single-stranded DNA and RNA
\cite{rosa:03,tkac:04}.
A chain-and-blob \cite{dunw:18} configuration 
can be observed where the blob at the end
fluctuates in shape and size due to thermal fluctuations
(see the supplementary video S1 for $L_p/L=0.2$ and $F_1=4$).
The corresponding PDF broadens while still displaying a single peak
which moves toward larger values of $R_g$.
At $F_1 \simeq 7$, the chain is stretched although, from time to time, the
final part can be still folded due to self-attraction
(see the supplementary video S2 for $L_p/L=0.2$ and $F_1=7$).
The PDF of $R_g$ exhibits two peaks corresponding
to  fully elongated and partially bent conformational states which are
stable for relatively long times to be clearly observed. This multi-peak
feature is similar to that observed for pulled semiflexible polymers
under poor-solvent condition
\cite{mare:04,ciep:04,rosa:06} and proteins subject to a  uniform flow
\cite{lema:03}. 
By further increasing the force, the polymer is completely elongated
with a narrow PDF of $R_g$ whose position shifts continuously to 
larger values
of $R_g$.
The relation (\ref{f}), observed once
$F_1 \gtrsim 10$, indicates that the chain behaves as a semiflexible
polymer under
strong force.
As a matter of comparison we report also the results
in the case
when self-attraction is negligible for a similar bending
rigidity
(see the data for $L_p/L=0.1$ and $\epsilon/(k_B T)=0.25$ in the left panel
of Figure \ref{fig:deficit}).
The behavior at small force values is different with 
the deficit length-ratio decaying as $F^{-1/2}$, which is
typical of semiflexible polymers without self-interaction.
By increasing the force, the dependence
(\ref{f}) is recovered
with the values of $\delta$ collapsing onto the ones
for $\epsilon/(k_B T)=2$.

\begin{figure}[ht]
\includegraphics*[width=.5\textwidth,angle=0]{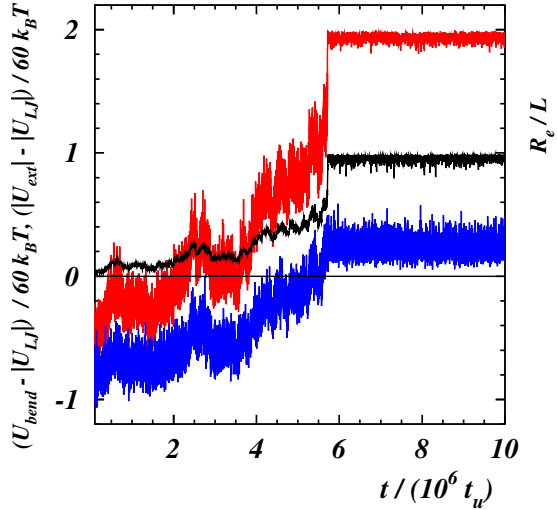}
\caption{Potential energy differences
$U_{bend} - |U_{LJ}|$ (blue line) and $|U_{ext}| - |U_{LJ}|$ (red line)
as functions of time in the case of the polymer with $L_p/L=0.8$ and
$\epsilon / (k_B T) = 2$ for $F_2=N F L_p / (k_B T)=200$.
The time behavior of the end-to-end distance $R_e$ is also
shown (black line).
\label{fig:ener}
}
\end{figure}

When the stiffness of the chain is such that a polymer exhibits
a double-stranded
conformation,
the mechanical response to the external force is different as it can be
seen in the right panel of Figure \ref{fig:deficit} where the average
deficit length-ratio
$\delta$ is plotted as a function of the
dimensionless force $F_2=N F L_p / (k_B T)$. Three regimes can be distinguished.
For values $F_2 \lesssim 10$ the two strands are aligned along the force
direction but there is no relative motion of the last bead with respect
to the first one, kept fixed in the origin, so that $\langle x_{N+1} \rangle \simeq 0$.
In this case the PDF of the gyration radius is narrow (see the curve
corresponding to the case with $L_p/L=0.8$ and $F_2=100$
in the right panel of Figure \ref{fig:pdfrg}).
When the force is increased, the strand, which is
not constrained to the origin, moves over the other part. This causes
a broadening of the PDF (see the curve with $F_2=120$).
A stronger force facilitates a larger sliding.
Due to the overall fluctuations of the polymer, the final bead does not
attain a fixed position relative to the first bead but can move back and forth
along the chain
(see the supplementary video S3 for $L_p/L=0.8$ and $F_2=150$).
The PDF is characterized by more peaks corresponding to different 
stable configurations assumed by the polymer in the same ensemble.
However, the final part of the chain cannot slide continuously
due to the finite
rigidity so that larger forces are required to unfold the polymer.
The time behavior of the energy terms  $(U_{bend} - |U_{LJ}|)$
and $(|U_{ext}| - |U_{LJ}|)$ is shown in Figure \ref{fig:ener}
in the case with $L_p/L=0.8$ for $F_2=200$.
The last bead can slide
when it occurs that $|U_{ext}| > |U_{LJ}|$, as witnessed by the increase
of the end-to-end distance $R_e$ also reported in the figure.
As $R_e$ gets continuously larger, 
$U_{bend}$ approaches $|U_{LJ}|$ and,
when $U_{bend}$ exceeds $|U_{LJ}|$, the polymer swells abruptly
signaling a ``first-order''-like phase transition.
(see the supplementary video S4 for $L_p/L=0.8$ and $F_2=200$).
Once the polymer is completely elongated, the PDF has again a single peak
whose position jumps discontinuously to a larger value.
The force required to unzip completely the polymer increases with the
bending rigidity and the transition from the folded state to the elongated one
becomes sharper, as in the case of the unzipping of double-stranded DNA
\cite{lube:02,tkac:04,kapr:09}.
When the polymer is completely unfolded, the values of $\delta$
for different bending rigidities lay on the same curve following the decay
(\ref{sf}) of semiflexible filaments, as it happens in the case of the stiffer
chain with $L_p/L=2$. 

\begin{figure}[H]
\includegraphics*[width=.9\textwidth,angle=0]{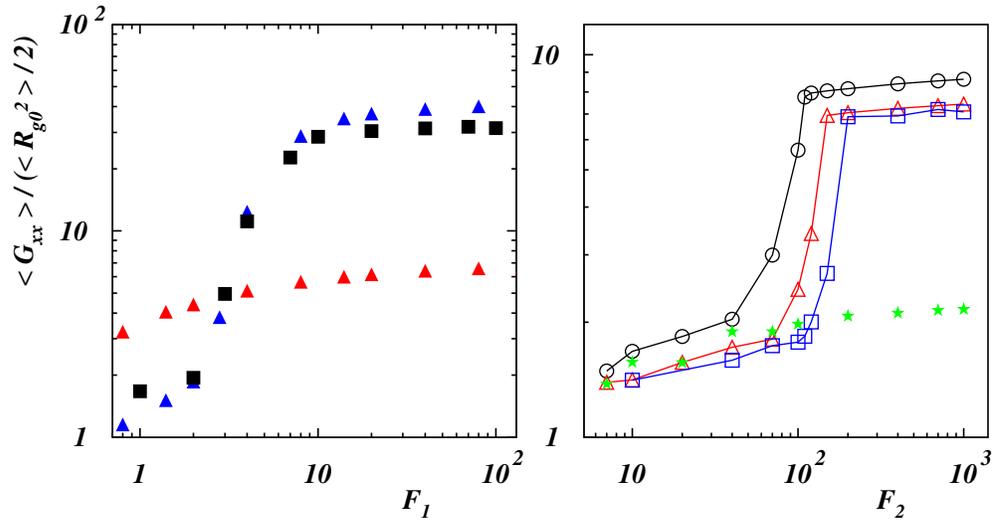}
\caption{(Left panel) Radius of gyration tensor component along the force direction
with respect to the equilibrium value
as a function of the dimensionless force $F_1=N F r_0 / (k_B T)$
for $L_p/L=0.1$ (blue filled triangles),
$0.2$ (black  filled squares) with $\epsilon / (k_B T) = 2$,
and for $L_p/L=0.1$ (red filled triangles)
with $\epsilon / (k_B T) = 0.25$.
(Right panel) Radius of gyration tensor component along the force direction
with respect to the equilibrium value
as a function of the dimensionless force $F_2=N F L_p / (k_B T)$
for $L_p/L=0.4$ (black open circles),
$0.6$ (red open triangles),
$0.8$ (blue open squares),
$2$ (green filled stars) with $\epsilon / (k_B T) = 2$.
\label{fig:gxx}
}
\end{figure}

\begin{figure}[H]
\includegraphics*[width=.9\textwidth,angle=0]{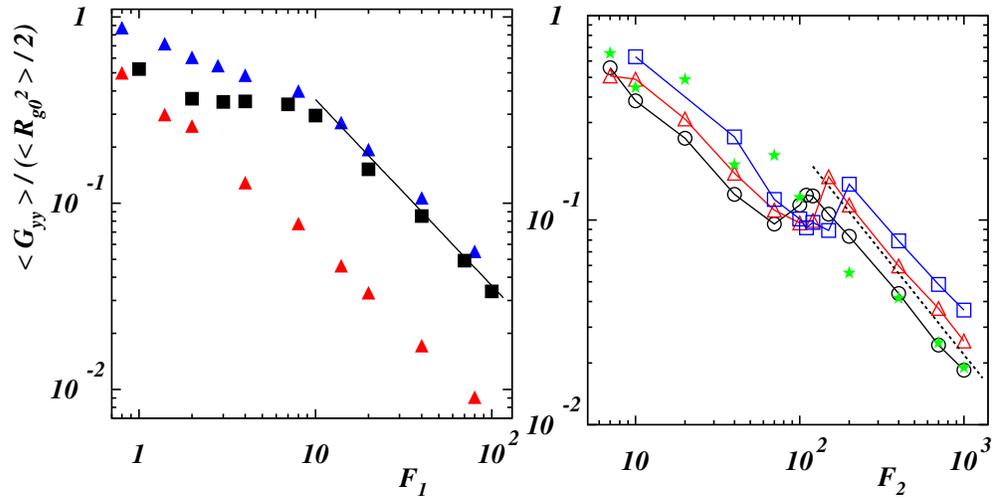}
\caption{(Left panel) Radius of gyration tensor component along the $y$-direction
with respect to the equilibrium value
as a function of the dimensionless force $F_1=N F r_0 / (k_B T)$
for $L_p/L=0.1$ (blue filled triangles),
$0.2$ (black  filled squares) with $\epsilon / (k_B T) = 2$,
and for $L_p/L=0.1$ (red filled triangles)
with $\epsilon / (k_B T) = 0.25$.
The full line has slope $-1$.
(Right panel) Radius of gyration tensor component along the $y$-direction
with respect to the equilibrium value
as a function of the dimensionless force $F_2=N F L_p / (k_B T)$
for $L_p/L=0.4$ (black open circles),
$0.6$ (red open triangles),
$0.8$ (blue open squares), $2$ (green filled stars)
with $\epsilon / (k_B T) = 2$.
The dashed line has slope $-1$.
\label{fig:gyy}
}
\end{figure}

Polymer deformation can be characterized in terms of the gyration tensor
(\ref{eq:tens}). The ratios 
$\langle G_{\alpha\alpha} \rangle/(\langle R_{g0}^2 \rangle/2)$
($\alpha \in \{x,y\}$) are presented in Figs.~\ref{fig:gxx},\ref{fig:gyy},
where  $\langle R_{g0}^2 \rangle$ is the mean-square value of the gyration radius
calculated at equilibrium.
For values of the bending rigidity $L_p/L \leq 0.2$
(see the left panel of Figure \ref{fig:gxx}), the behavior
is similar and the polymer
is smoothly deformed in the force direction as long as the blob is unwound
($1 \lesssim F_1 \lesssim 10$).
Once the chain has been disentangled ($F_1 >10$),
the deformation reaches a value which does not change significantly with the
force. Due to the initial compact structure, the ratio of deformation is
considerably larger with respect to the case with negligible
self-interaction which is also shown in the left panel 
of Figure \ref{fig:gxx} for  $L_p/L = 0.1$. In the right panel of the same figure
the deformation is shown as a function of the dimensionless force $F_2$
when the polymer has a double-stranded initial
configuration. Initially, in case of $\langle x_{N+1} \rangle \simeq 0$, 
the force slightly elongates the chain with respect to
the equilibrium case. As soon as the last bead starts to 
slide over the filament,
the deformation increases rapidly with steepness depending 
on the bending rigidity. Finally, it reaches a constant value
when the polymer is fully elongated along the force direction.
The smaller relative deformation corresponds
to the more stiff polymer whose initial configuration is a stiff hairpin
(see Figure \ref{fig:conf0} (d)). When the bending rigidity is such that no
closed structure can form ($L_p/L=2$), the chain is smoothly elongated over
the whole range of explored forces with a final value sensibly smaller than
the one corresponding to initially double-stranded chains.

Along the $y$-direction, normal to the force, the relative
deformation
diminishes as a function of the dimensionless force $F_1$ when $L_p/L \leq
0.2$ (left panel of  Figure \ref{fig:gyy}).
As long as the chain maintains a compact structure, the decrease is weak
while it becomes steeper when the polymer is open under the action of the
external driving. At values $F_1 > 10$, data collapse and
a power-law with dependence $F_1^{-1}$ can be observed.
When self-interaction is negligible, 
the behavior is similar but the deformation is much 
smaller due to the lack of a compact-like initial structure.   
More interesting appears to be what happens for the range of stiffness
corresponding to double-stranded conformations. The initial values
of $\langle G_{yy} \rangle/(\langle R_{g0}^2 \rangle/2)$
decrease, due to the stretching of the two strands,
with a similar trend. When the folded strands open, an overshoot can be
observed that is due to the larger fluctuations of the chain.
The deformation  
then follows a power-law decay with dependence $F_2^{-1}$.
The data for the  initially stretched polymer ($L_p/L=2$)
show a similar
behavior without the aforementioned overshoot.

%%%%%%%%%%%%%%%%%%%%%%%%%%%%%%%%%%%%%%%%%%
\section{Discussion and Conclusions} \label{sec:conclusions}

The dynamical and conformational properties of
semiflexible polymers under poor-solvent condition in a
uniform force field have been numerically studied. The chain has been anchored
at one end, confined in two dimensions and placed in contact with a Brownian
heat bath implemented by the stochastic version of the multiparticle collision
dynamics.

The equilibrium conformation depends both on the stiffness and on
the self-interaction
strength. When the
self-attraction energy is smaller compared to the thermal energy,
the chain behaves as a semiflexible filament. In the opposite limit
of strong
mutual attraction, different configurations are obtained.
At low bending rigidity the polymer assumes a compact structure.
By increasing the stiffness, patterns of folded bundles
emerge where the number of
strands reduces as the chain becomes more rigid.
A larger number of polymer beads, with respect to the value here considered,
would promote folded conformations with more strands
as observed for three-dimensional semiflexible polymers \cite{seat:13}.
Finally, rod-like conformations are
recovered for high values of the rigidity.

The mechanical response to the action of the external force depends
on the initial equilibrium structure. For small bending rigidity the
compact structure is continuously unwound and stretched by the force.
On the other hand, when the polymer consists of two facing strands,
a ``first-order''-like phase transition
is observed from the folded to the stiff conformation.
These behaviors are highlighted in the force-extension relations
as well as at the probability distribution functions of the gyration radius.
The deformation of the radius of gyration with respect to the equilibrium value
along the direction normal to the force is found to decay as the
inverse of the applied force.

Although hydrodynamics interactions have been neglected in this
investigation, it is known that such interactions are not essential
in the case of semiflexible polymers since only logarithmic corrections
are expected \cite{harn:96}. Therefore, the present results
also describe the behavior of a self-attractive semiflexible
polymer placed in a uniform flow field as long as the chain follows the fluid
flow. We hope that this study will stimulate theoretical studies and
experimental investigations to confirm the outlined phenomenology.

%%%%%%%%%%%%%%%%%%%%%%%%%%%%%%%%%%%%%%%%%%
\vspace{6pt} 

%%%%%%%%%%%%%%%%%%%%%%%%%%%%%%%%%%%%%%%%%%
%% optional
\supplementary{Video S1: Animation of polymer stretching for $L_p/L=0.2$ and $\epsilon/(k_B T)=2$
  with $F_1=4$; Video S2: Animation of polymer stretching for $L_p/L=0.2$ and $\epsilon/(k_B T)=2$
  with $F_1=7$; Video S3: Animation of polymer stretching for $L_p/L=0.8$ and $\epsilon/(k_B T)=2$
  with $F_2=150$; Video S4: Animation of polymer stretching for $L_p/L=0.28$ and $\epsilon/(k_B T)=2$
  with $F_2=200$.}

%%%%%%%%%%%%%%%%%%%%%%%%%%%%%%%%%%%%%%%%%%
\funding{This research was funded by MIUR project number PRIN 2020/PFCXPE.}

\dataavailability{Data are available upon reasonable request.} 

\acknowledgments{A. L. wishes to thank R. G. Winkler for critically reading the manuscript
and A. Rosa for useful discussions.
This work was performed under the auspices of GNFM-INdAM.}

\conflictsofinterest{The author declares no conflict of interest.} 

%%%%%%%%%%%%%%%%%%%%%%%%%%%%%%%%%%%%%%%%%%
\begin{adjustwidth}{-\extralength}{0cm}

\reftitle{References}
\bibliography{polymer_bib}

%%%%%%%%%%%%%%%%%%%%%%%%%%%%%%%%%%%%%%%%%%
\end{adjustwidth}
\end{document}